\documentclass[runningheads]{llncs}
\usepackage[T1]{fontenc}
\usepackage{graphicx}
\usepackage{svg}
\usepackage{subcaption}
\begin{document}
\title{\texttt{MIND}: AI Co-Scientist for Material Research}
\author{Geonhee Ahn\inst{\star}, Donghyun Lee\thanks{Equal contribution}, Hayoung Doo, \\
Jonggeol Na\inst{\star\star}, Hyunsoo Cho\inst{\star\star} \and Sookyung Kim\thanks{Corresponding authors} }
\authorrunning{G. Ahn et al.}
\institute{Institute for Multiscale Matter and Systems (IMMS), Ewha Womans University, Seoul 03760, Republic of Korea \\\email{\{jgna, chohyunsoo, sookim\}@ewha.ac.kr}\\
}
\maketitle

\begin{abstract}
Large language models (LLMs) have enabled agentic AI systems for scientific discovery, but most approaches remain limited to text-based reasoning without automated experimental verification. 
We propose \texttt{MIND}, an LLM-driven framework for automated hypothesis validation in materials research. 
\texttt{MIND} organizes the scientific discovery process into hypothesis refinement, experimentation, and debate-based validation within a multi-agent pipeline. 
For experimental verification, the system integrates Machine Learning Interatomic Potentials, particularly \textit{SevenNet-Omni}, enabling scalable \textit{in-silico} experiments. 
We also provide a web-based user interface for automated hypothesis testing. The modular design allows additional experimental modules to be integrated, making the framework adaptable to broader scientific workflows.
The code is available at: \url{https://github.com/IMMS-Ewha/MIND}, and a demonstration video at: \url{https://youtu.be/lqiFe1OQzN4}.


\end{abstract}
\section{Introduction}

Recent advances in LLMs have enabled agentic AI systems for scientific discovery. Across scientific domains, LLM-based approaches have been explored for literature mining%~\cite{wu2025automated}
, hypothesis generation~\cite{bazgir2025agentichypothesis}, and research planning~\cite{gottweis2025towards}. However, most systems remain limited to \textit{text-based reasoning} and do not perform experiments required for validating scientific hypothesis. This limits the realization of fully automated closed-loop scientific discovery.
In this work, we propose \texttt{MIND} (Materials INference \& Discovery), an LLM-driven framework for automated hypothesis validation in materials research. \texttt{MIND} structures the scientific discovery process into three stages: pre-experiment, experiment, and discussion, implemented as a multi-agent pipeline. After discussion, if the evidence is sufficient, the system generates a validation report as the final output; otherwise, it revises the hypothesis and returns to the \textit{Pre-Experiment} stage, enabling iterative refinement and validation.
A key challenge in closed-loop discovery is experimental verification. While integration with physical laboratories remains challenging, computational simulations provide a practical alternative. We therefore employ Machine Learning Interatomic Potentials (MLIPs) as the experimentation module. In particular, we use \textit{SevenNet-Omni}\cite{kim2026optimizing}, a foundation MLIP model capable of predicting diverse material properties across many materials systems, enabling scalable \textit{in-silico} experiments.
To support practical use, we provide an interactive user interface that allows materials scientists to easily run automated hypothesis testing. The architecture of \texttt{MIND} is modular and extensible, allowing additional experimental modules to be integrated into the experimentation stage and enabling adaptation to broader scientific discovery pipelines.
We evaluate \texttt{MIND} on hypothesis concerning three target material properties—energetic, mechanical, and structural. On a benchmark of domain-expert-designed hypothesis, \texttt{MIND} achieves overall $75.0\%$ average accuracy.  A user study with experimental and computational materials scientists also shows strong satisfaction with the system’s usability. 
%and research support capability.

%\noindent\textbf{Contributions.}
%We propose \texttt{MIND}, an agentic AI framework for \textit{fully automated closed-loop hypothesis validation} in materials research. The system integrates hypothesis refinement, MLIP-based experimentation, and evidence-driven multi-agent debate into a unified pipeline that enables experimental verification of hypothesis. The framework is modular and extensible, allowing additional experimental modules to be incorporated for broader scientific discovery workflows.

%%%%%%%%%%%%%%%%%%%%%%%%%%%%%%%%%%%%%%%%
\begin{figure}[t]
    \centering
    \includegraphics[width=\textwidth]
    {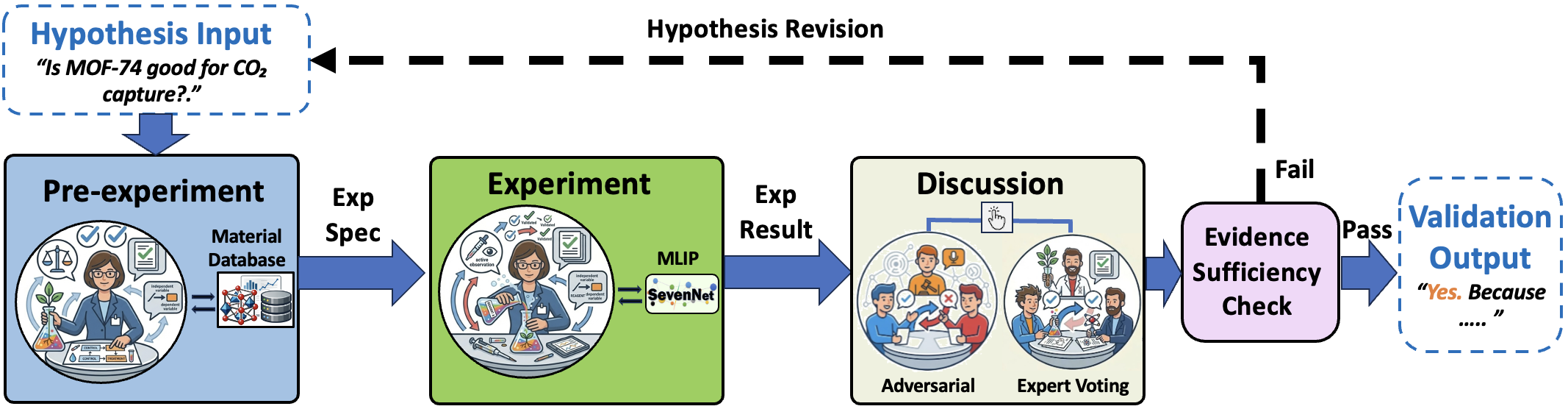}
    \vspace{-12pt}
    \caption{Overview of automated hypothesis validation process in \texttt{MIND}.}
    \vspace{-12pt}
    \label{fig1}
\end{figure}
%%%%%%%%%%%%%%%%%%%%%%%%%%%%%%%%%%%%%%%%

\vspace{-6pt}

%%%%%%%%%%%%%%%%%%%%%%%%%%%%%%%%%%%%%%%%
\begin{figure}[t]
    \centering
    \includegraphics[width=\textwidth]{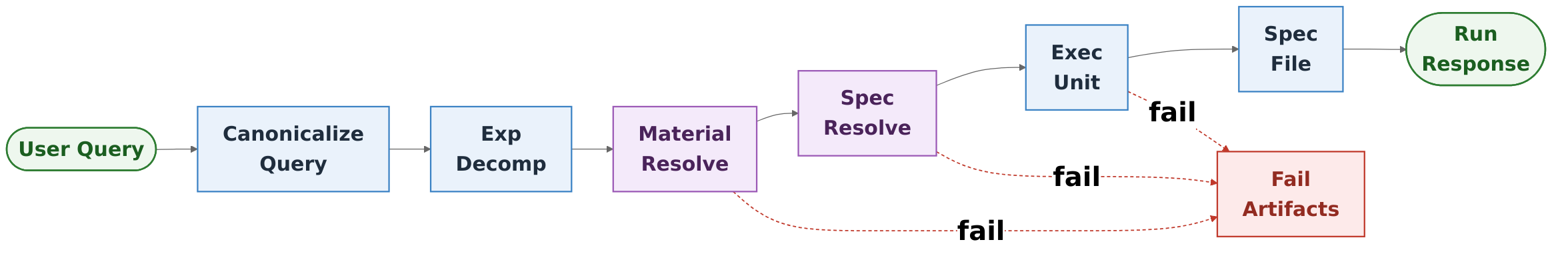}
    \vspace{-18pt}
    \caption{Pre-experimental stage workflow.}
    \vspace{-18pt}
    \label{fig2}
\end{figure}
%%%%%%%%%%%%%%%%%%%%%%%%%%%%%%%%%%%%%%%%

\section{\texttt{MIND} Workflow}

Inspired by the scientific research process, \texttt{MIND} performs automated hypothesis validation in three stages: \textbf{Pre-Experiment}, \textbf{Experiment}, and \textbf{Discussion} (Fig.~\ref{fig1}). 
Each stage is a module in a \textit{LangGraph}-based multi-agent pipeline.
%This design enables structured reasoning, automated computational experimentation, and evidence-based hypothesis evaluation.

\vspace{0.2cm}\noindent
\textit{Pre-Experiment.}
Given a scientific hypothesis, the pre-experiment stage transforms natural-language user input into structured specifications for MLIP calculations. The system first canonicalizes the hypothesis, extracting user intent, the research questions to be addressed, and the target materials. Subsequently, the material-resolution module retrieves structural data for each target material in CIF format from reference databases, establishing the structural provenance required for reproducible simulation. The spec-resolution module then resolves simulation parameters, including calculator configuration (model, precision, device, seed) and task parameters (optimizer, force convergence threshold, maximum steps, cell relaxation). The resolved information is assembled into discrete execution units, each corresponding to a unique material–trial pair, and serialized as schema-validated JSON specification files that serve as direct input contracts for the subsequent MLIP calculation stage. If any execution unit fails — due to an unresolvable material, unresolvable simulation parameters, or a schema validation error — a structured failure report is recorded in place of a valid specification, allowing the remaining units to proceed independently.
The detailed workflow of pre-experiment stage is shown in Figure~\ref{fig2}.

\vspace{0.2cm}\noindent
\textit{Experiment.}
The experiment stage performs automated \textit{in-silico} MLIPs, specifically \textit{SevenNet-Omni}. 
Given the experimental specification produced in the previous stage, the framework executes simulations to predict target material properties. 
Through the Claude Model Context Protocol (MCP), the system connects to a remote compute server where SevenNet-Omni simulations are executed. 
The experimental JSON specification is transmitted to the remote server, which performs the simulations and returns a structured JSON output containing predicted properties and simulation results. 
These outputs serve as quantitative evidence for evaluating the hypothesis.

\vspace{0.2cm}\noindent
\textit{Discussion.}
In the final stage, multiple agents analyze the experimental results and assess whether the evidence supports the hypothesis. 
We consider two discussion strategies. \textbf{(1) Adversarial Discussion} employs three agents—\emph{supporter}, \emph{skeptic}, and \emph{judge}. 
The supporter argues in favor of the hypothesis, while the skeptic challenges its validity. 
After several rounds of debate, the judge synthesizes the arguments and produces the final decision.  \textbf{(2) Expert Voting} instead uses multiple scientist agents that analyze the experimental evidence collaboratively and independently vote on the hypothesis validity, with the final decision determined by majority voting.
Based on the discussion outcome and experimental evidence, the system evaluates whether the available evidence is sufficient to validate the hypothesis. 
If the evidence is sufficient, the framework generates a validation report as the final output. 
Otherwise, the system revises the hypothesis and returns to the \textit{Pre-Experiment} stage, enabling iterative hypothesis refinement and validation.

\begin{figure}[t]
\vspace{-6pt}
\centering
\includegraphics[width=0.95\linewidth]{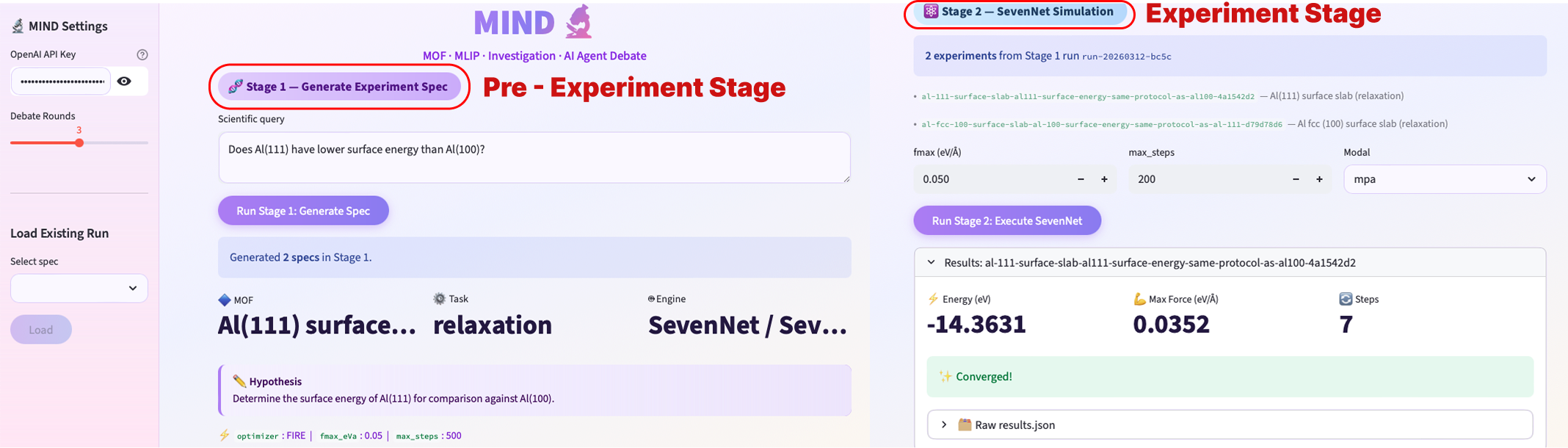}
\vspace{-6pt}
\includegraphics[width=0.95\linewidth]{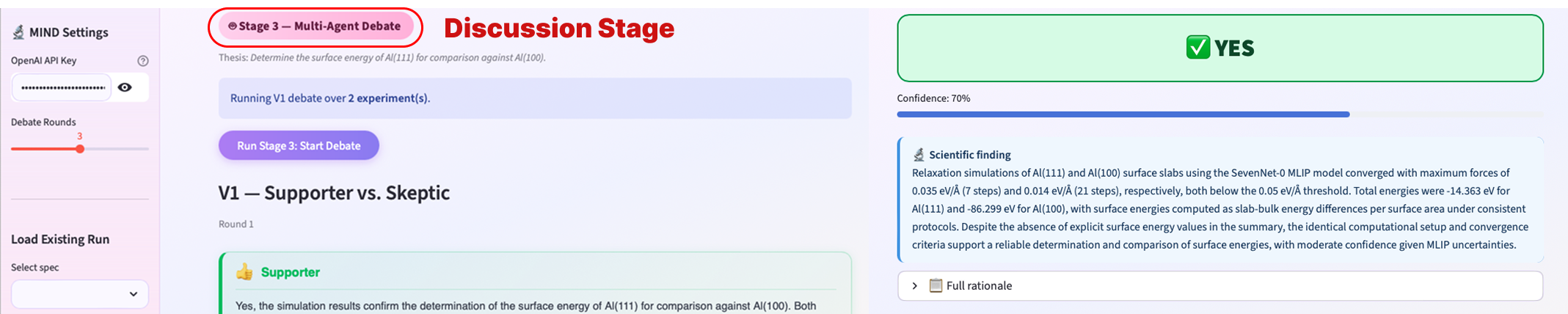}
%\vspace{-5pt}
\caption{The interactive user interface to submit hypothesis and visualize results.}
\vspace{-12pt}
\label{fig:mind_pipeline}
\end{figure}

\vspace{0.2cm}\noindent
\textit{User Interface.}
To enable practical use by researchers, we also develop a web-based interface implemented with \textit{Streamlit}. (Shown in Fig~\ref{fig:mind_pipeline})
The interface allows users to submit hypothesis, monitor experimental progress, and visualize debate results, enabling interactive exploration of the \texttt{MIND} framework.
%%%%%%%%%%%%%%%%%%%%%%%%%%%%%%%%%%%%%%%%
% TODO Figure 2. \MIND UI Component 
%%%%%%%%%%%%%%%%%%%%%%%%%%%%%%%%%%%%%%%%
\section{System Evaluation}

We evaluate the \texttt{MIND} framework from two perspectives: 
(1) \textbf{system accuracy}, measuring how accurately the framework validates scientific hypothesis, and 
(2) \textbf{system utility}, assessing its usefulness for real materials research.

\vspace{0.1cm}\noindent
\textit{System Accuracy.}
To evaluate hypothesis validation accuracy, we construct an MLIP-expert–curated benchmark where each hypothesis is a materials science claim with simulation-verifiable ground truth. 
The benchmark spans three property categories—energetic, mechanical, and structural—and each task is formulated as a binary (\textit{yes}/\textit{no}) verification problem.
Across 28 hypothesis, \texttt{MIND} correctly validates 21 cases, achieving an overall accuracy of $\textbf{75.0\%}$, with category accuracies of $70\%$ (energetic), $75\%$ (structural), and $100\%$ (mechanical). 
Among the correct predictions, 8 cases required iterative hypothesis refinement through additional experiment cycles, highlighting the benefit of closed-loop validation.
On average, \texttt{MIND} verifies a hypothesis in \textbf{5 minutes}. Compared to the typical \textbf{3–6 hour} human research loop using SevenNet-omni, this represents a \textbf{36–72$\times$} speedup in hypothesis verification.

\vspace{0.1cm}\noindent
\textit{System Utility.}
To evaluate practical usefulness, we conduct a user study with 26 materials scientists, including both experimental and computational researchers. Participants are presented with outputs from \texttt{MIND}, including the input hypothesis, predicted validation result, and reasoning trace. They evaluate the system along three dimensions: 
(1) \textbf{scientific validity of the results}, 
(2) \textbf{reasoning transparency}, and 
(3) \textbf{research usefulness}, using a 7-point Likert scale.
The average scores are 5.76 for scientific validity, 5.78 for reasoning transparency, and 5.88.
Across all dimensions, most participants rate the system above the neutral threshold (i.e, 4 point), suggesting that \texttt{MIND} provides meaningful support for hypothesis validation in materials research.

%%%%%%%%%%%%%%%%%%%%%%%%%%%%%%%%%%%%%%%%
% TODO Table.  Results
%%%%%%%%%%%%%%%%%%%%%%%%%%%%%%%%%%%%%%%%
%\section{Conclusion and Future Work}
%We propose \texttt{MIND}, an agentic AI framework for \textit{automated closed-loop hypothesis validation} in materials research that integrates hypothesis refinement, MLIP-based experimentation, and evidence-driven multi-agent debate in a modular pipeline. 
%As future work, we plan to extend \texttt{MIND} to support more complex hypothesis and integrate physical laboratory experiments for automated real-world validation.

 \bibliographystyle{splncs04}
 \bibliography{reference}

\end{document}